\documentclass[preprint,saps,prl,onecolumn,superscriptaddress]{revtex4}
\usepackage{graphicx}
\usepackage{times}
\usepackage{amsmath}
\usepackage{textcomp}

\begin{document}

\title{ Experimental realization of SQUIDs with topological insulator junctions}

\author{M. Veldhorst}
\affiliation{Faculty of Science and Technology and MESA+ Institute for Nanotechnology, University of Twente, 7500 AE Enschede, The Netherlands}
\author{C.G. Molenaar}
\affiliation{Faculty of Science and Technology and MESA+ Institute for Nanotechnology, University of Twente, 7500 AE Enschede, The Netherlands}
\author{X.L. Wang}
\affiliation{Institute for Superconducting and Electronic Materials, University of Wollongong, Wollongong, NSW, 2522, Australia}
\author{H. Hilgenkamp}
\altaffiliation[Also at ]{Leiden Institute of Physics, Leiden University, P.O. Box 9506, 2300 RA Leiden, The Netherlands}
\affiliation{Faculty of Science and Technology and MESA+ Institute for Nanotechnology, University of Twente, 7500 AE Enschede, The Netherlands}
\author{A. Brinkman}
\affiliation{Faculty of Science and Technology and MESA+ Institute for Nanotechnology, University of Twente, 7500 AE Enschede, The Netherlands}\date{\today}

\begin{abstract}
We demonstrate topological insulator (Bi$_2$Te$_3$) dc SQUIDs, based on superconducting Nb leads coupled to nano-fabricated Nb-Bi$_2$Te$_3$-Nb Josephson junctions. The high reproducibility and controllability of the fabrication process allows the creation of dc SQUIDs with parameters that are in agreement with design values. Clear critical current modulation of both the junctions and the SQUID with applied magnetic fields have been observed. We show that the SQUIDs have a periodicity in the voltage-flux characteristic of  $\Phi_0$, of relevance to the ongoing pursuit of realizing interferometers for the detection of Majorana fermions in superconductor- topological insulator structures.
\end{abstract}

\maketitle

A three-dimensional topological insulator is an insulator in the bulk, but has conducting surface states that can be described by means of a Dirac cone in which the spin is locked to the electron momentum \cite{Zhang2006, Fu2007, Zhang2009N, Qi2009, Hsieh2008, Chen2009, Hsieh2009N, Peng2009, Cheng2010, Zhang2009P}. Electrons with opposite spin have opposite momenta, which suppresses backscattering, rendering these materials interesting for low power electronic devices. Combining the helical Dirac fermions with a superconductor \cite{Fu2008, Nilsson2008, Tanaka2009} may lead to the artificial creation of the elusive Majorana fermion \cite{Majorana1937}. In the search for the Majorana fermion, efforts have been made to contact a topological insulator (TI) to a superconductor (S). Supercurrents in S-TI-S junctions have been reported \cite{Dumin2011, Sacepe2011, Veldhorst2011}. The combined evidence for a ballistic Josephson supercurrent and the presence of topological surface states, show that topological Josephson junctions can be made, despite the presence of a conductivity shunt through the bulk of the TI crystal \cite{Veldhorst2011}.

In this Letter, we demonstrate dc SQUIDs consisting of Josephson junctions with topological insulator surface states as barrier layer. These interference devices potentially serve as a basis to detect the Majorana fermion. Majorana bound states may appear in the vortex of a topological superconductor and at S-TI interfaces \cite{Fu2008, Tanaka2009}. In the latter, time reversal symmetry in the topological insulator needs to be broken, for example by incorporating a magnetic insulator layer or by applying an external magnetic field. Intriguing devices are proposed to identify the appearance of this exotic particle \cite{Fu2008, Nilsson2008, Tanaka2009, Beenakker2011}. Since the Majorana fermion is charge neutral and is a zero energy state, most proposals rely upon quantum interference devices. For example, a current-phase relationship with a $4\pi$ periodicity \cite{Fu2008} might result from the interplay between the Majorana fermion and a superconductor. We therefore  study the current-phase relationship of the topological dc SQUID.

\begin{figure}[t]
	\centering 
		\includegraphics[width=0.5\textwidth]{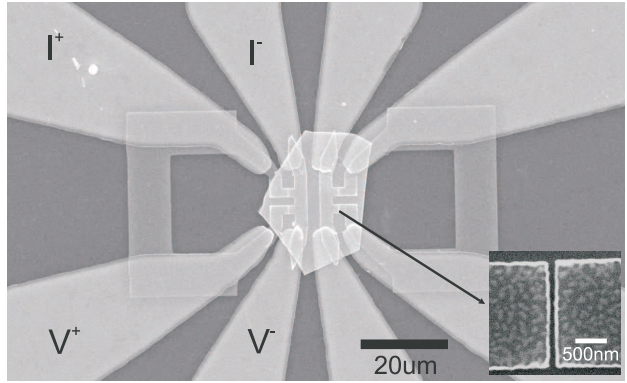}
		\caption{Scanning Electron Microscopy image of topological insulator nano dc SQUIDs. Two SQUIDs with length $l$=120nm are designed on an exfoliated Bi$_2$Te$_3$ (200nm height) flake. The two Nb arms are differently sized resulting in an asymmetric SQUID. Nb is sputter deposited and defined by e-beam lithography. Inset shows a Josephson junction.}
		\label{fig:1}
		\vspace{-15pt} 
\end{figure}

\begin{figure}[t]
	\centering 
		\includegraphics[width=0.3\textwidth]{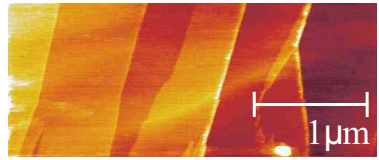}
		\caption{ Atomic force microscopy image of an exfoliated Bi$_2$Te$_3$ surface. The step edges are 1.0~nm high, corresponding to the Bi$_2$Te$_3$ quintuple unit cell. These nanometer flat surfaces span an area up to 50 $\times$ 50~\textmu m$^2$.  }
		\label{fig:2}
\end{figure}

Figure \ref{fig:1} shows two superconductor Nb - topological insulator Bi$_2$Te$_3$ dc SQUIDs fabricated on one Bi$_2$Te$_3$ flake. We have fabricated polycrystalline Bi$_2$Te$_3$ samples with a common $c$-axis orientation using the Czochralski method as described elsewhere \cite{Li2010}. Using mechanical exfoliation, Bi$_2$Te$_3$ flakes ranging from 30 nm to 1 \textmu m are transferred to a Si substrate. The Bi$_2$Te$_3$ flakes are  smooth on the nm scale over areas of several  \textmu m$^2$ scale. Figure \ref{fig:2} shows an Atomic Force Microscopy image of the surface of a typical Bi$_2$Te$_3$ flake, revealing the 1.0 nm quintuple unit cell layers of Bi$_2$Te$_3$. After exfoliation, a superconducting ($T_C=9K$) Nb-layer (200nm) is sputter deposited with a 5nm Pd layer deposited \textit{in situ} on top to protect the Nb against oxidation. Electrodes are defined by photolithography. The Nb on top of the Bi$_2$Te$_3$ flake makes a strong superconducting contact with the Nb on the substrate. Finally, nanojunctions are defined by lift-off e-beam lithography and sputter deposition of Nb. The substrate is slightly conducting ($\rho=5\Omega\textrm{cm}$) at room temperature to increase the resolution of e-beam lithography, but is completely insulating at low temperatures. Prior to deposition, \textit{in situ} Ar-ion etching is performed in order to make transparent contacts. It is found that Ar etching roughens the Bi$_2$Te$_3$ surface, probably by preferential etching, but 2 minute etching at 50 eV leaves a Bi$_2$Te$_3$ surface with 1-2 nm roughness and a transparent contact. Throughout the fabrication process the Bi$_2$Te$_3$ surface in between the Nb leads has only been covered by resist, leaving the surface of the junction barrier layer unaffected from Ar-ion etching and deposition steps.  Two separate electrodes connected by Nb over the Bi$_2$Te$_3$ flake, for example the positive current and voltage leads, have a superconducting contact with a critical current exceeding 30 mA, ensuring a large supercurrent through all layers.

We have designed two SQUIDs on one Bi$_2$Te$_3$ flake (200nm height), see Fig. \ref{fig:1}. The nanojunctions are 2 \textmu m wide and have a length $l= 120$ nm. The inductance ratio between the two arms is about $\lambda=0.2$. In a square washer approximation \cite{Ketchen85} the total effective area of the SQUID is 920 \textmu m$^2$, and the estimated inductance 46 pH. Both SQUIDs showed similar behavior, but in the rest of the paper we focus on the left SQUID of Fig. 1. 

\begin{figure}[t]
	\centering 
		\includegraphics[width=0.5\textwidth]{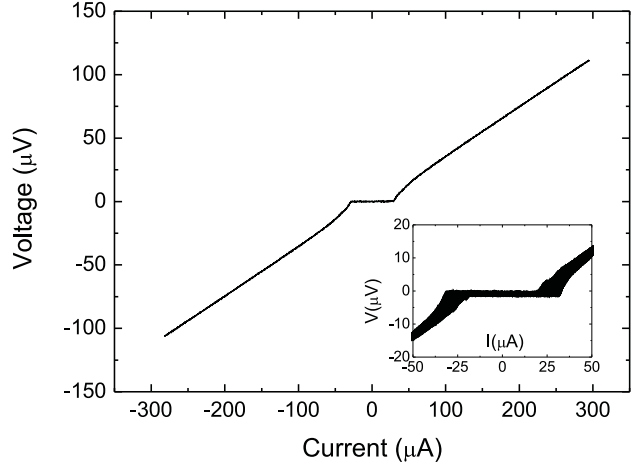}
		\caption{$IV$-curve of the topological dc SQUID with atypical critical current of 30\textmu A. The junctions are 2\textmu m wide and have an electrode separation of 120 nm. The intrinsic bulk shunt has reduced the $I_CR_N$ product to 10 \textmu V. Inset shows the $IV$-curve under magnetic field modulation.}
		\label{fig:3}
		\vspace{-15pt} 
\end{figure}

Below 6 K, the superconducting proximity effect induces a Josephson supercurrent through the junctions and at 1.4 K the dc SQUID has a critical current of 30 \textmu A, as shown in figure \ref{fig:3}. This results in a 2D critical current per width of the junction of 7.5 A/m, which is within 10$\%$ of the individually measured critical current density of junctions on different flakes. The SQUID had a critical current constant within 10$\%$ over three cooldowns running over several weeks. Due to shunting of the junctions results we find a relative low $I_CR_N=10$ \textmu V. Ballistic junctions with $\xi_0 \approx$ 80nm, $T_C=6 K$ and $l$=120 nm, have an estimated $I_cR_N \approx 130-260$\textmu$V$, but the shunt due to the bulk conductance reduces the characteristic voltage to a few percent of the expected value \cite{Veldhorst2011}, consistent with the observed 10 \textmu V. In order to obtain higher $I_cR_N$ values, electrical gating or chemical substitution, e.g.  Bi$_2$Se$_3$Te$_3$ \cite{Jia2011}, could be used. The current-voltage characteristics are intrinsically non-hysteretic because of the low capacitance of the lateral geometry of the junctions and the high transparency of the interfaces.

\begin{figure*}
	\centering 
		\includegraphics[width=1\textwidth]{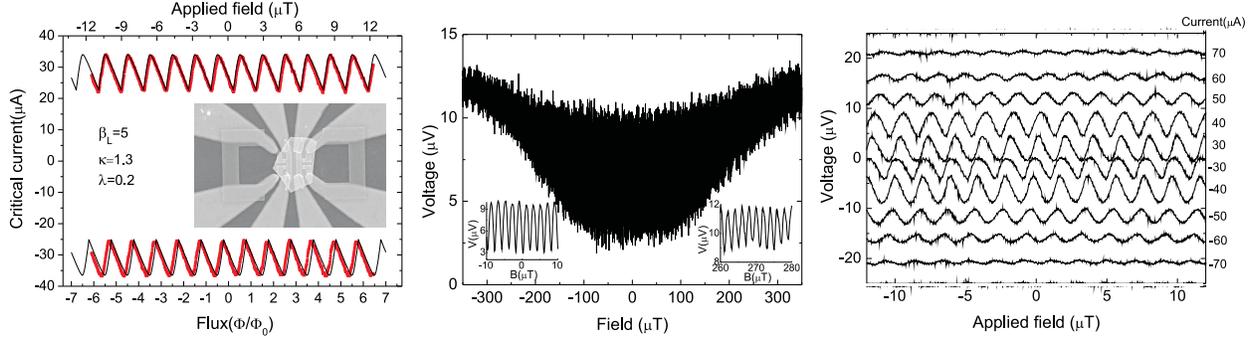}
				\vspace{-15pt} 
		\caption{Voltage modulation the dc SQUID. (a) The critical current of the SQUID can be fitted (thin line) by a model that accounts for the asymmetry of the design. (b) The critical current of the individual junctions is suppressed at 350\textmu T (Fraunhofer pattern). SQUID oscillations are still observable. Inset shows the SQUID oscillations at small fields (left) and large fields (right) (c) The V-$\phi$ relationship reveals an asymmetric response as well as a shift for increasing bias currents. The solid lines represent equicurrents for $\pm$ 30-70 \textmu A.}
		\label{fig:4}
		\vspace{-15pt} 
\end{figure*}

Applying an external magnetic field causes oscillations of the superconducting critical currents due to interference of the two arms, shown in \ref{fig:4}. This modulation of $I_c$ unequivocally demonstrates the correct operation of the Nb-Bi$_2$Te$_3$ SQUID. The voltage modulation is strongest at a bias of approximately 40 \textmu A, which corresponds to a current close to the critical current of the SQUID. The critical current modulation is $\sim 35\%$ of the total critical current. Using a simple model based on a sinusoidal current-phase relation of the junctions, and taking inductance and possible asymmetries into account, we can fit the field dependence with high accuracy. The asymmetrical inductance causes an asymmetry in the SQUID current-phase relationship, see Fig. \ref{fig:4}. We also observe a shift in the current-phase relationship when increasing the bias current. The two arms of the SQUID have different inductances, which results in a growing field threading the SQUID for increasing current. The 0.75 $\Phi_0$ shift in the V-$\phi$ relation at 90\textmu A bias can be accounted for by a 10pH inductance difference between the two arms. This corresponds to the fitted values for the total inductance of the SQUID, $L = \frac{\Phi_0}{2\pi}\frac{\beta_L}{I_c} = 55\textrm{pH}$ and the asymmetry factor $\lambda = 0.2$, resulting in an inductance difference of 11pH. At the optimum bias, the sensitivity of the SQUID is $\textrm{15 \textmu V}/\Phi_0$. In the limit $\beta_L(=5) \gg 1$, the sensitivity of a SQUID can be estimated by $R/L$ \cite{Tesche1977} which yields $\textrm{12 \textmu V}/\Phi_0$.

As expected, the SQUID modulation frequency is much larger than that of the individual junctions, see Fig. \ref{fig:4}. A SQUID oscillation corresponds to 1.9 \textmu T, while the critical current of the junctions is suppressed at 350 \textmu T. This corresponds to about 180 SQUID oscillations in a junction oscillation, which is slightly lower than expected by comparing the enclosed areas. However, for the enclosed area of a junction, the Josephson penetration depth and flux focussing has to be included, which increases the junction effective area \cite{Veldhorst2011}.

Since the supercurrent is carried by topological surface states \cite{Veldhorst2011} we can study the  current-phase relationship of toplogical junctions by means of our dc SQUID. In a dc SQUID with topological insulator surface states as interlayers, the phase difference over the junctions due to a given field is equal to a standard SQUID. However, the presence of Majorana fermions can cause a $I_c=I_0 \sin(\phi/2)$ junction current-phase relation, resulting in a  $4\pi$ periodic dependence of the junctions \cite{Fu2008}. To test the periodicity of the SQUID, the magnetic field from the coil is calibrated with a Hall sensor and the effective area of the junction, 920\textmu m$^2$, is estimated using the square washer approximation \cite{Ketchen85}. A  $2\pi$ periodic dependence within 7 $\%$ accuracy is calculated, enough to exclude $4\pi$ periodicity. The absence of a  $4\pi$ periodic dependence might result from the absence of a ferromagnetic insulator (breaking time reversal symmetry), or due to relaxation to equilibrium states as was predicted theoretically \cite{Fu2009, Badiane2011}.

In conclusion, we have demonstrated the Nb-Bi$_2$Te$_3$ SQUIDs. As shown previously, the supercurrent is carried by the topological surface states, thereby allowing the study of the current-phase relationship of superconductor- topological insulator structures. From the V-$\phi$ characteristics we deduced a  $2\pi$ current-phase relationship. The high reproducibility allows to study the noise properties of these devices in future experiments and to eventually include magnetic insulators in order to break time reversal symmetry to create Majorana bound states.

This work is supported by the Netherlands Organization for Scientific Research (NWO) through VIDI and VICI grants, by the Dutch FOM foundation, and by the Australian Research Council through a Discovery project.

\end{document}